\documentclass[showpacs,twocolumn,preprintnumbers,amsmath,amssymb,a4paper]{revtex4-1}

\usepackage{amsfonts}
\usepackage{amsmath}
\usepackage{amssymb}
\usepackage{times}
\usepackage{amstext}
\usepackage{amsthm}
\usepackage{graphicx, subfigure}
\usepackage{dcolumn}
\usepackage{bm}
\usepackage{natbib}
\usepackage[usenames,dvipsnames]{xcolor}

\begin{document}

\begin{center}

{\LARGE How crude oil prices shape the global division of labour}

\vspace{0.3cm}
Francesco Picciolo$^{1,\ast}$, Andreas Papandreou$^2$, Klaus Hubacek$^3$ \& Franco Ruzzenenti$^{4,5}$

$^{1}$Department of Chemistry, University of Siena, Via Aldo Moro 2, Siena, 53100, Italy

$^{2}$Department of Economy, University of Athens, 8 Pesmazoglou Street, Athens, 10559, Greece

$^{3}$Department of Geographical Sciences, University of Maryland, CollegePark, Maryland, USA

$^{4}$Department of Economics and Statistics, University of Siena, Siena, Italy

$^{5}$ Institute of Sociology, Jagiellonian University, Krakow, Poland

$^\ast$To whom correspondence should be addressed; E-mail:  picciolo.fr@gmail.com

\end{center}

\begin{abstract}
Our work sheds new light on the role of oil prices in shaping the world economy by investigating flows of goods and services through global value chains between 1960 and 2011, by means of Markov Chain and network analysis. We show that over that time period the international division of labor and trade patterns are tightly linked to the price of oil. We demonstrate that this correlation does not depend on the balance of payments nor on the nominal value of trade or trade agreements; it is instead linked to the way the Global Value Chains (GVCs) shape global trade. Our study suggests that transport played an important structural role in shaping GVCs.
\end{abstract}


\section*{Introduction}

In the aftermath of the oil crisis of the early 1970s, the relationship between oil prices and economic growth became a focal point of the scientific discourse and public debate. In 1983, James Hamilton published an influential article showing that an oil price increase had preceded all but one recession in the United States since the end of World II\cite{hamilton}. Since then, a large number of empirical studies have looked into the connection between oil prices and real economic growth and frequently found a significant negative correlation\cite{jones,allsopp}.
 The importance of this link between oil price and economic growth was less clear after the second oil shock\cite{jones04,hooker,mork,alvarez}. Recent studies, with more refined statistical tools and price specifications, have accomplished in restoring the between the oil price and economic growth\cite{hamilton11,naccache,papa,ola,cologni}.
 There is now a general consensus on the notion that this did not cease but it has become more complex in terms of direction (anticyclical and procyclical), typology of shocks (demand or supply) and lag patterns\cite{jones04,naccache,ola,barsky}. 
This line of research tried to explain this tight relationship, given that the cost of energy is but only a small part of GDP\cite{barsky} but satisfactory explanations have remained elusive\cite{jones,jones04,barsky,park08oil}. Interestingly, this research on the oil price and economy economic indicators seems to have entirely ignored the transport sector, which is heavily reliant on refined crude oil products, and its role in shaping the global division of labour.
 In the post-war period, world trade grew at a faster pace than world GDP\cite{hummels07}. According to recent studies on globalization, the remarkably high rate was propelled by a dramatic decline in international transport costs\cite{hummels07,coe07,anderson04,feenstra97}. Perhaps, the notion that trade grew amid globalization because of transport should not come as a surprise. What is more surprising, but is closely related or even a corollary, is the fact, that intermediate and capital goods, in the last decades, grew faster than final products and now account for the largest part of trade in OECD countries\cite{miroudot}. While, in the aftermath of World War II international trade mainly concerned final products, the second wave of globalization (since the late 1980s) extended to intermediate products and capital goods, and the integration of factors’ market as another important effect\cite{baldwin}. This process led to the fragmentation of production internationally\cite{hummels01}. Disregarding the transport sector, most of the scholars focused their attention on other factors in order to explain the fragmentation of the global value chain, like the pursuit of cheap labour or more favourable environmental legislations\cite{baldwin,hummels01,levinson}.  Amador and Cabral recently suggested that the strong increase of trade associated with the development of the global value chains (GVCs) in the 1990s coincides with a period of low oil prices, although admitting that there is little empirical evidence linking these two factors\cite{amador14}. These findings suggests the importance of assessing the impact of oil shocks in an internationally integrated system rather than on a national base only. Furthermore, the new issues posed by climate change demand a deeper understanding of the nexus between energy consumption and the global economic structure. Our study addresses the connection between oil price and the global economy, by means of network theory and Markov chain theory, with the aim of understanding how the GVCs expanded and shrank following price changes in crude oil, between 1960 and 2011. In contrast to previous analyses, that progressed by refining price specifications and statistical methods, we observed the correlation of economy with crude oil price (Brent), but we changed the macro-economic variables under investigation. We first applied network theory to trade imbalance and bilateral trade to understand how these two global measures of trade are linked to the oil price. These two quantities are thereby used to introduce the cycling index that builds on Markov chain analysis to assess the amount of value that is conserved across direct and indirect relationships in trade. With this measure, we looked at the share of cyclical value –the share of value that returns to the starting point, along different paths in the world trade network.

\section*{Analysis and Results}

\subsection{Balance of trade per country: trade (im)balance}
The balance of trade is the difference in value of exports and imports (see methods). Many have viewed the existence of large current account imbalances between large economies as a possible cause of the financial crisis\cite{carrasco,rebucci}. There is growing evidence that current account (im)balances are correlated to oil prices worldwide\cite{rebucci}. The reason for this correlation lies in the burden placed on imports (or exports, for oil exporting countries) by energy commodities, but also in monetary policies aimed at regulating inflation (which is correlated to oil price)\cite{rebucci}.
The analysis has been performed on a yearly basis between 1960 and 2011, on aggregate trade flows (total import/export for every country) of all the reporting countries in the world. Data are taken from Gleditsch’s\cite{gled} and BACI datasets\cite{zignago}. The fluctuation in the balance of trade and the variation of the adjusted crude oil price are slightly, yet statistically significantly, negatively correlated: the linear correlation coefficient is -0.32 (see tab.1). It is noteworthy that in a network where flows tend to be balanced at every vertex, the matrix tends to be symmetrical (which means that entries in the upper triangular matrix mirror those in the lower triangular one)\cite{rec}. In other words, symmetric weights (flows) between every pair of vertices is statistically the simplest way to balance ingoing and outgoing flows at every vertex. A local  symmetry (exports equals imports) tends to produce a global symmetry (export from $i$ to $j$ equal export from $j$ to $i$). We thus expect that the balance of bilateral trade in the world trade web (WTW) to be negatively correlated to oil prices because we observe a negative correlation of the balance of trade locally. It should be noted though, that this is just statistical relationship, obtained by imposing the local balance as a constraint in the null model\cite{rec}.

\subsection{Balance of bilateral trade: trade reciprocity}
A global measure for evaluating the balance of trade between every pair of countries is the weighted reciprocity\cite{rec}. Reciprocity is a first-order property, meaning that it concerns the direct relationship of nodes with the nearest topological neighbours (one link-length). Reciprocity has proven to be an helpful measure in understanding the effects of the structure on dynamic processes, explaining patterns of growth in out-of-equilibrium networks\cite{vinko11,diego05}, and starting to evaluate higher order properties\cite{zamora,vinko09,milo,motifs,camacho}. The reciprocity for weighted networks, $r_w$, is defined as follows:

\begin{equation}
r^w= \frac{\sum_i \sum_j \min[w_{ij},w_{ji}]}{\sum_i \sum_j {w_{ij}}}.
\label{eq:reciprocity}
\end{equation}

If all flows are perfectly reciprocated/balanced then $r^w=1$. If  they are unidirectional $r^w=0$. The correlation of $r^w$ with oil price is -0.70. As the oil price increases (decreases), reciprocity decreases (increases). This result shows that changes in the international bilateral trade structure and the oil price are more intimately and inversely linked than expected by observing only global volumes or distances of trade\cite{hummels07,coe07,anderson04}. These results deliver two important pieces of information:1) given that the correlation of the oil price with reciprocity is higher than the correlation with imbalance, the nexus between oil price and reciprocity (bilateral trade balance) cannot be reduced to the correlation between oil price and imbalance, despite the expected symmetrical effect (see above); 2) this tight correlation is not explainable with trading agreements, as it is ubiquitous\cite{francosymmetryII}, nor with a general abatement of barriers\cite{hummels07}, as it has a discontinuous trend in time (see SI).
Can we link reciprocity to some structural effect in production, like the development of GVCs? 
As previously stated, reciprocity is a first-order property, though, unless we assume that production chains involve only two sites in a row, GVCs should be pertinent to higher-order properties of networks. In the SI we show (fig. 4), with a heuristic model based on three countries, one product and two factors, how reciprocity increases when the production chain expands, involving second order properties of the network (neighbours of neighbours, or indirect relationship). Within this simple model, the shift from a single-country production chain to a multi-country production chain will always increase the reciprocity of the network, independently on the distribution and share of the total volume traded. We thus expect that shifting the production sites abroad increases the reciprocity of the network. However, to check this hypothesis we must extend the analysis to longer paths of the productive chain, involving more than three nodes (like in the heuristic model) and, most importantly, encompassing indirect relationships.

\subsection{Markov chain analysis: cyclic paths of value in the world economy.}
The largest share of trade in the world economy involves inputs to production (raw materials, intermediate and capital goods)\cite{miroudot}. In the modern economy countries import these production inputs and export final products or intermediate goods that are further processed elsewhere often involving numerous production stages in many different countries. At every step in the global production chain value added is embodied\cite{amador14}. We are interested in detecting the share of traded value that is conserved throughout the stages of GVCs, or the initial value at the beginning of a global production chain, like mass particles in ecological networks that are conserved throughout every stage of a food chain\cite{finn} (see methods for a detailed description of the concept). Thus, we want to assess the share of trade that is cyclical – that runs a cyclic path, along paths of a given length $S$, in the WTW. By means of Markov chain theory, we can statistically evaluate the probability of an "elementary" trade, i.e. the amount of value embodied in raw materials or intermediates that is conserved in a product, going from country $i$ to country $j$ and return to the initial country throughout all the possible direct and indirect trade relationships\cite{markov1,markov2}. For example, Fig.1 illustrates all the possible cyclic paths that a particle (a unit value of trade, in our case) can follow from country $i$ within the first 4 steps (i.e. stages of production).

\begin{figure}
\begin{center}
{
\includegraphics[width=0.45\textwidth]{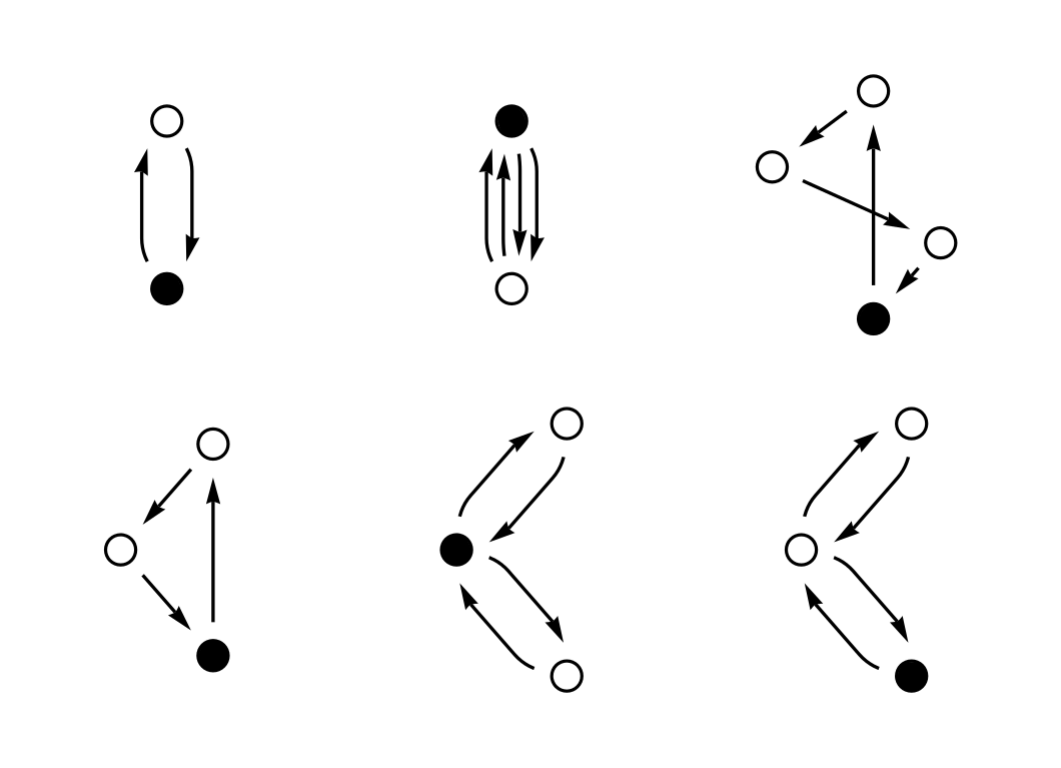}
}
\end{center}
\caption{Cyclic paths up to four steps. The possible cyclic paths that a trade can follow on a network within 4 steps. Node $i$ (represented by the black dot) is the starting and ending country of each path. $\Gamma^{(S)}$ evaluates the share of the total trade that follow a cyclic path of length up to $S$.}
\label{fig1}
\end{figure}

In order to assess the share of cyclical value over the total value traded, we need a normalized measure, like the previous measures of imbalance and reciprocity. We indicate with the cycling index $\Gamma^{(S)}$ the share of trade that comes back to the starting country in $S$ steps with $S = 2,…,\infty$   (see methods). In Fig. 2 trends of the cyclical quantity are shown for the WTW, for $S = 2, 3, 4, \infty$.
We observe that the percentage of cyclical value inside the network and oil price are negatively correlated, at various degrees. The lower the number of steps (i.e. the shorter the production chain) taken into account while evaluating $\Gamma^{(S)}$ the higher the negative correlation. The correlation from $\Gamma^{(2)}$ to $\Gamma^{(\infty)}$ goes from -0.85 to -0.62, exhibiting a much higher score compared to imbalance and reciprocity (see tab.1).

\section*{Discussion}
\subsection{Digging into the tight relationship between cycling and oil price.} One may argue that a general increase in exports and imports driven by global GDP growth, would inevitably lead to raise the share of mutual trade. In this view, the correlation between cycling and oil price could be brought back to the GDP-oil price nexus. Moreover, it is also possible that oil prices inflating commodity prices and automatically increasing nominal trade flows, could explain this tight correlation. In this latter case, the relationship would boil down to the link between inflation and oil price. Thirdly, it is possible that the correlation between oil price and cycling is a multiplicative effect (re-spending effect of petrodollars) of the correlation between imbalance and oil price, linking oil producing with oil consuming countries. This is conceivable on the notion that oil exporters tend to import goods and services from oil importers. In summary, all three may be included in the notion of first order properties of the network contrary to cycling that is a second order property.
We can show, using a Null Model as a benchmark that the negative correlation between the cyclical value inside of the network and oil price does not depend on first order properties (direct relationships), but depends on higher-order properties of the network (indirect relationships). The chosen null model is the Weighted Directed Configuration Model (WDCM). The WDCM is a well-known and utilized null model that preserves the quantity of trade and the distribution of trade relationship of each country\cite{rec}. We performed the analysis of the null model and we evaluated the correlation of the expected cyclical value with the oil price (see methods). The correlation between the expected $\Gamma^{(S)}$'s and the oil price are significantly lower than the observed ones. They range from -0.32 to -0.38. Notably by using the WDCM we preserve: 1) the nominal value of all the exports/imports sequence; 2) the global value of trade; 3) the difference between import and export (balance of payments) at every vertex. The comparison of expected correlation from the Null Model with the observed correlation assures us that the latter is not due to the global volume traded, nor to the nominal values of trading relationships or the distribution of trade imbalance: it depends on the specific architecture of trade flows globally, which can only be understood by the way GVCs unfold.
In order to understand the underlying process, it is also instructive to observe the trend of cycling indexes of different path-length over time. In Fig.2 and Fig.3, the 2-steps length peak of the early 1990s is lower than the one of the late 1960s, meaning that the higher orders contribution to the cycling has become more and more important, following the decreasing trend of the 1970s. If we look at the trend of the normalized values of the two, three and four steps cycling compared to oil price (Fig.3), it is evident that until the mid-1970s the three degree of cycling overlap, while between the first and the last oil crisis (2008) the curves of the three and four step cycling stand above that of the two step cycling. The longer-than-two step cycling also displays higher growth rates until the peak of the 1990s, suggesting that the ongoing globalization was characterized by longer paths of cycling. Longer cycles in the value chain probably underline the growing share of intermediate goods in trade and a process that led to a more interdependent global economy (see SI).

\begin{figure}
\begin{center}
{
\includegraphics[width=0.45\textwidth]{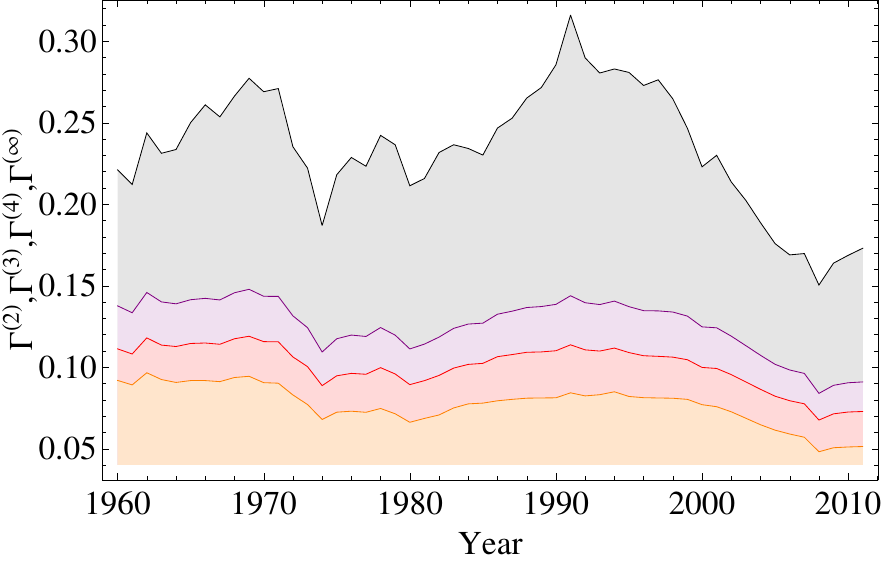}
}
\end{center}
\caption{The trend of the cyclical flow index, $\Gamma^{(S)}$, of the World Trade Web calculated in eq.(5) from 1960 to 2011. From lighter to darker color: $\Gamma^{(2)}$ (orange line), $\Gamma^{(3)}$ (red line), $\Gamma^{(4)}$ (purple line), and $\Gamma^{(\infty)}$ (black line).}
\label{fig2}
\end{figure}

\subsection{The second wave of globalization and the role of transport}
We hypothesize that the transmission mechanism behind the correlation between oil price and GVCs, statistically measured by cycling, is the transport sector. Oil prices impact on transport costs, making international outsourcing more or less profitable\cite{francorebound,transport}. If this hypothesis is correct, we expect oil prices to impact on cycling, by influencing the length of GVCs. We tested Granger causality, between oil price and the three network measures here considered (imbalance, reciprocity and 2-step-cycling), in both directions, with one lag specification and a significance level of 5\%\cite{granger}. Only in the case of cycling we can reject the null hypothesis that oil price does not Granger cause 2-step-cycling.
The Granger test, however, indicates that causation runs in both directions. Indeed, the most important information brought about Granger causality test is that 2-step-cycling is cointegrated with oil price, meaning that these two variables follow the same trend, pointing to a long-term dynamics. By enlarging the scope of the analysis from first order properties of the network to higher order properties, we obtained not only a higher correlation to oil, but also a cointegration, meaning that the analysis through the cycling quantity are able to detect and capture the global changes of the pattern of WTW. It thus seems plausible to believe that transport is the nexus between GVCs and oil price, as we expect changes in the structure of production to occur in the long-term. However, there are many other factors that influence the global division of labour that would need to be included to better explain this relationship. What seems somewhat surprising, given the declining costs of transport, is the fact that the process of international integration peaked in the 1990s and declined in the 2000s. Indeed, the second half of the 1990s witnessed the onset of the second wave of globalization\cite{baldwin}.
This apparent contradiction is probably explained by observing single-country cycling (the portion of cycling passing through a single vertex, see fig. 4). During the 2000s the single-cycling of fast-growing economies, like China, increased dramatically, climbing the ranking of the World's economies. While the cycling index of developed economies remained constant, China showed a rapid growth in the share of cyclical value of its trade, meaning that a large part of the GVCs began passing through this country. Some economies like China attracted a significant portion of the cycling value across the World, concentrating the flows into a few hubs. The decrease in cycling globally is consistent with the emergence of hubs, which are topologically like stars (see fig.4 in SI). It is plausible that outsourcing initially, from the late 1980s, was propelled by road transport, over medium distances and involving many countries (inflating global cycling) whereas in the second stage, from the 1990s, was fuelled by air and cargo shipping, concerning longer distances and few hubs (reducing global cycling). According to IEA, international cargo shipping, mostly in non-OECD countries, displayed the highest growth rates among the different transport modes between 1990-2000\cite{transport}.  It is noteworthy that the first major shift coincides with a sudden leap in efficiency in road transports triggered by the oil crisis\cite{francorebound}, whereas the second change followed a dramatic decrease in the international transports costs, both in air and cargo shipping, following the introduction of a more efficient aircraft fleet and the containers system\cite{hummels07}. To investigate this hypothesis, we assessed how the role of distances in shaping the cyclical value has changed over time.  If we trim the network at different distance thresholds (removing all the links placed between a couple of vertices above a certain distance value) and calculate the cycling, we observe that there have been a bifurcation between short and long distance cycling in the early 1990s (see fig.6 in SI). While before the 1990, the cycling indexes trimmed at different distances were moving coherently, after the 1990s the share of short-distance cycling index (less than 2500 km) began to decline compared the share of long-distance cycling index, which kept growing (more than 2500 km). This seems to indicate that the GVC's shifted from a regional domain (clusters of neighbouring countries) to a global (oversea relationships) domain.

\begin{table}[h]
\centering
\begin{tabular}{| c || c | c |}
\hline
     & R & p-value   \\ \hline\hline
$ b_t$ & -0.32 &  0.02 \\  \hline
$ r^w$ & -0.70 &  $<10^3$ \\  \hline   
$\Gamma^{(2)}$ & -0.85 & $<10^3$ \\  \hline
   $\Gamma^{(3)}$ & -0.83 & $<10^3$ \\  \hline
   $\Gamma^{(4)}$ & -0.82 & $<10^3$ \\  \hline
   $\Gamma^{(\infty)}$ &  -0.62 & $<10^3$ \\  \hline
 \end{tabular}
\caption{The linear correlation index between $b_t$, $r^w$, $\Gamma^{(S)}$ (for $S \in \{2,3,4,\infty \}$) and the crude oil price are reported, in the third column the respective p-values are shown ($95\%$ confidence).}
    
\label{tab:corr}
\end{table}

\section*{Concluding Remarks}

\begin{figure}

\label{fig3}

\begin{center}
{
\includegraphics[width=0.45\textwidth]{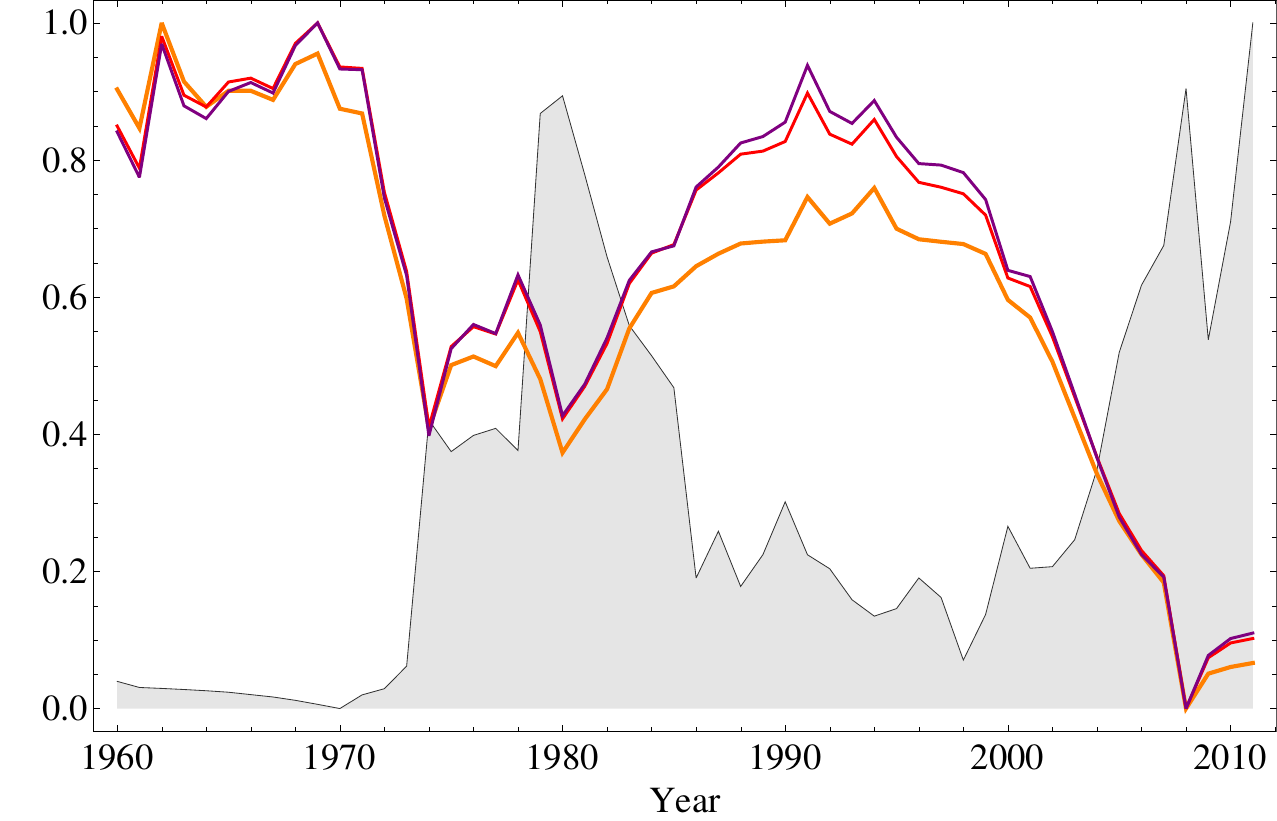}
}

\caption{The normalized crude oil price (black solid line with subtended grey area) is shown. The trend of the normalized value of  $\Gamma^{(2)}$ (black solid line), $\Gamma{(3)}$ (dashed line), $\Gamma^{(4)}$ (dotted line) are also shown. The linear normalization is done based on the maximum and minimum value reached by each quantity. The final scaled values lie between 0 and 1.}
\end{center}
\end{figure}

\begin{figure}
\begin{center}
{
\includegraphics[width=0.45\textwidth]{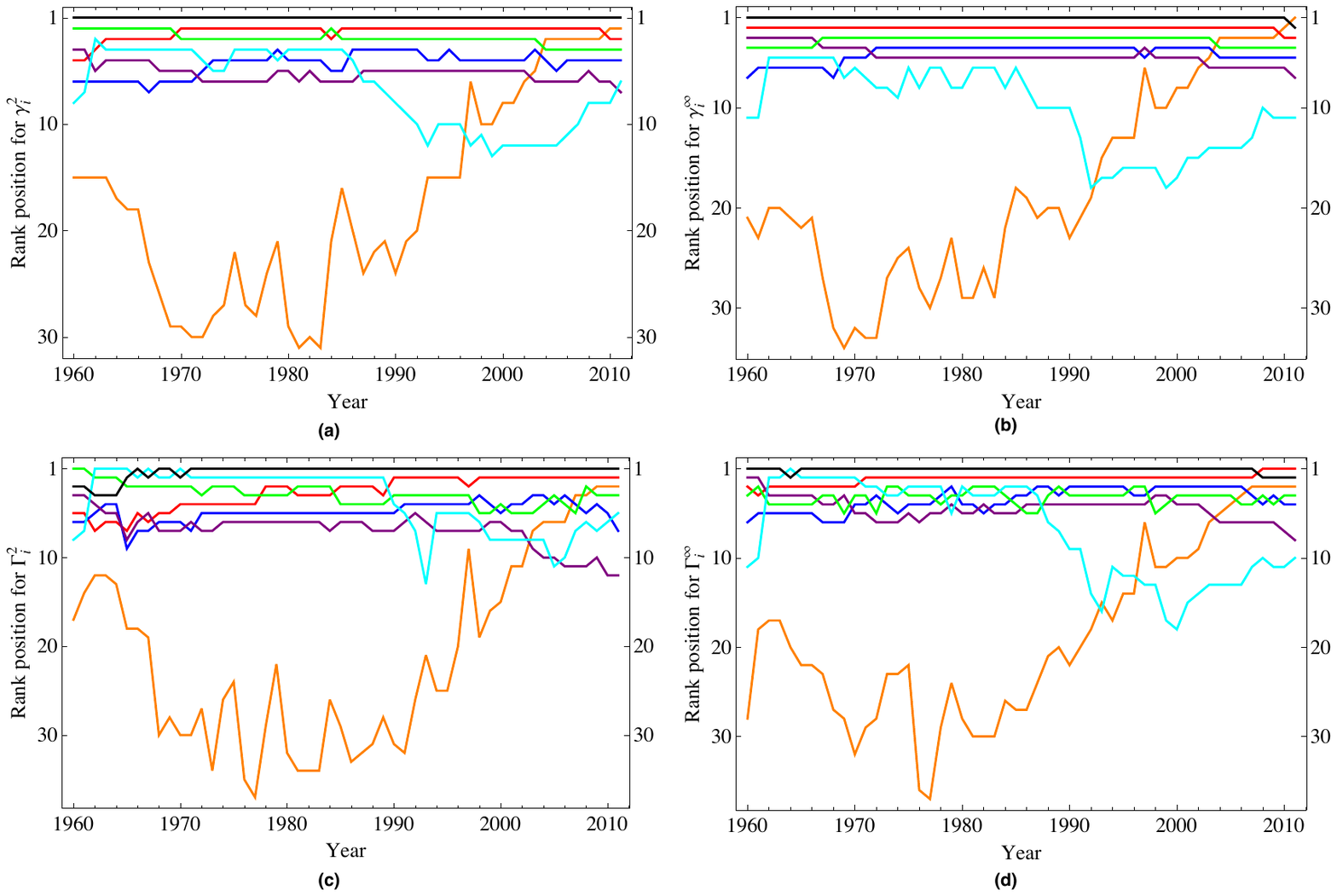}
}
\end{center}
\caption{The ranking position for seven countries according to the value of $\gamma_i^{(2)}$ (panel a) and $\gamma_i^{(\infty)}$  (panel b), $\Gamma_i^{(2)}$  (panel c) and $\Gamma_i^{(\infty)}$ (panel d). The data range covers from 1960 to 2011. The seven countries are the following: China (orange), Germany (red), France (bleu), Great Britain (purple), Japan (green), Russia (cyan), USA (black). We chose these countries for their interesting patterns in order to reveal how the variables $\gamma_i^{(S)}$ and $\Gamma_i^{(S)}$ can explain structural changes in economy. Russia position is always higher in the relative than in the absolute rank, and for $S=2$ than for $S=\infty$. Moreover we can observe a decline in both starting in late 80s and a rise since around 2000. Results seem in line with the USSR dissolution, explained by a Russian chain value made up substantially by bilateral exchanges with other former Soviet Republics. A remark on the performance of China and Germany: the former has been on the rise in the last decades and we point out the pervasiveness of its chain value in absolute value reaching the top of the rank for $\gamma_i^{(\infty)}$ in the last years of our analysis. In the same period, Germany overtakes USA share of cyclic trade ($S=\infty$ ), as emerges from panel d.}
\label{fig2}
\end{figure}

Global trade, in the age of the second globalization, has entangled national economies, by interconnecting production sites internationally, in a fashion that is still underappreciated\cite{baldwin,amador14} [20,23]. This historical process of vertical integration has been producing global value chains (GVCs) wherein goods travelling across countries augment their value at every stage of production\cite{amador14}. We investigated this cyclic path of value across countries by means of network theory and Markov chain theory\cite{rec,markov1,markov2}. In this way, we show that over a longer time period the oil price has a striking correlation with the structure of trade globally. A correlation that increases with the scope of the analysis, from first order properties of network (one link distance), to higher order properties. The worldwide sum of country trade imbalances show a weak correlation of -0.32. The correlation increases by engulfing bilateral relationships between countries on a global scale (reciprocity, -0.70). Finally, the highest correlation, up to -0.85, is observed when we involve more complex pattern at the global level. By means of statistical mechanics of networks (exponential random graphs) we were able to demonstrate that this remarkable correlation can only be explained with higher-than-one order properties of the network, indicating that GVCs and structure of trade are intimately linked to oil price.
We hypothesize that this tight relationship points to the role of transports in determining, in the long run, the extent and the way production sites connect internationally.
By  looking closely at the single-country cycling index (fig. 4) and by dissecting the global cycling according to different distance thresholds (fig. 1 in SI), we individuated two structural breaks and two phases of the second wave of globalization (the “second unbundling”). The first, which started in the 1980s and peak in the 1990s, was featured by  shorter distance and increasing cycling, the second, that probably lasted until the economic crisis, was featured by longer distances, declining global cycling and increasing cycling of China (star-like structure).
Our results suggest that the transmission mechanism between oil price and economic growth lays not only in the labour or retail markets (via inflation), but also, more profoundly, in the structure of production globally.  Furthermore, in a more general perspective, our results indicate that the production structure could be approached as an energy system, constrained by energy efficiency in the transport sector. This view of the economic system builds on the work of scholars like Ayres\cite{ayres}, for approaching growth as a product of energy efficiency, but also on fundamental advances in the study of allometric scaling, aimed at explaining the structure and size of many biological processes as the result of general features of efficient transportation networks\cite{banavar}. Nevertheless, in order to establish a clear causation among factors, further research should tackle two aspects more in depth: 1) the dynamic process between input/output matrices of national economies and sectoral trade; 2) the evolution of energy efficiency and costs of transports globally, in the long run. The authors of this article believe this is crucial for understanding the role of oil in the present economic system, given that this is not yet a fungible source of energy in the transport sector, and for paving the way for a prosperous economy freed from fossil fuels.

\appendix
\section{Materials and methods}

\subsection{Description of the dataset.}
In the following sections, a brief description of the analysed networks is given. We analyse the series of yearly bilateral data on exports and imports among world countries from the Gleditsch's database\cite{gled}, from 1960 to 1997. From 1997 to 2011, we employed international trade data provided by the BACI database\cite{zignago}. All data are in millions of current U.S. dollars and all they are freely available.
Also the crude oil price data are freely available\cite{oil}.

\subsection{Evaluating the balance of trade.}
In network approach the World trade web is described by matrix$W$, where each node $i$ is a country and $w_{ij}$ is the trade from country $i$ to country $j$, we can evaluate trade imbalance as follows:

\begin{equation}
b^t= \frac{\sum_i \min[s^{in}_i , s^{out}_i]}{\sum_i \sum_j {w_{ij}}}
\label{eq:baloftrade}
\end{equation}

where $s^{in}_i=\sum_j   w_{ji}$  is the in-strength (total import) of node $i$ and $s^{out}_i=\sum_j   w_{ij}$  is the out-strength (total export) of node $i$. In doing so, we do not distinguish between trade surplus or deficit (positive or negative value), instead we calculate only the amount of trade that is balanced between imports and exports for every country (i.e. the minimum of total imports and exports); $b^t=1$ if exports and import are equal for all the countries, $0 \geq b_t>1$ otherwise.

\subsection{Evaluating patterns of traded value.}
We want to assess the probability of a percentage of trade to be in a country $j$, starting from $i$, after a number of production steps. Markov chain analysis allows us to statistically determine this. The probability is calculated according to the trade relationships between country in WTW reported by\cite{gled,zignago}.

\emph{Calculating transition probabilities.} In probability theory a Markov chain is a stochastic process defined on a discrete state space satisfying the Markov property. It is a set of random variables, representing the evolution of a certain system, without memory: each actual state of the system just depends on the previous one. The changes of states are called transitions and the Markov chain can be described by a Markov matrix, $M$, whose elements $m_{ij}$ represent the probabilities from a state $i$ to a state $j$ (transition probabilities or transition rules). Given the memory-less nature of a Markov chain it is not possible to predict the state of a system, in a given time t, but it is possible to predict its statistical properties. In this paper the transition probabilities are set according to the elements of the matrix $W$ that describes the World Trade Web, i.e. the trade network. A link, $w_{ij}$ , is assigned to any import/export relationship between two countries, from $i$ to $j$ , where $w$ is the volume of the trade as provided by datasets\cite{gled,zignago}. 
In what follow, we explain how the transition rules are assigned. In general, given a system of $N$ nodes and a $N \times N$ matrix representing their interactions, the matrix is said balanced if if $s^{out}_i=s^{in}_i,\:\forall\:i \in \{1, \dots , N \}$,  i.e., the sum of each column is equal to the sum of its related row. Systems described by a balanced matrix can be considered isolated: nodes perfectly balance their total in- and out-flows themselves, without needing any further exchange with the outside. However, most of the network representing real systems are not balanced.
This means that $s^{out}_i \neq s^{in}_i$ for at least one $i$ such that $ i \in \{1, \dots , N \}$. In these systems three different sets of nodes can be identified: the set of vertices with $s^{out}_i=s^{in}_i$, the set of vertices with $s^{out}_i>s^{in}_i$ and the set of vertices with $s^{out}_i<s^{in}_i$.
In the first group, the total ingoing and outgoing fluxes of all nodes are balanced. In the second set, each vertex needs an extra incoming weight in order to balance in and out strengths. In this case, in order to balance the nodes in this set, we introduce an additional node/vertex, called \emph{source} and labeled with 0, providing extra-ingoing fluxes to all nodes in the set.
The new link between the source and each node in the set will have the following weight: $w_{0i}=s^{out}_i - s^{in}_{i}$. Similarly, a new vertex is introduced for the nodes in the third group in order to balance their in-strengths. It is called \emph{sink} and labeled $N+1$. The new links between this latter and the nodes in the set will have weights equal to: $w_{i (N+1)}=s^{in}_i-s^{out}_{i}$. 
The vertices source and sink play the role of the internal economies of the countries. Trade surplus and trade deficit are absorbed by internal economies.
For each node $i$ with $i \in \{1,\dots,N \}$ the \emph{total outgoing flow}, $v_{i}$, is given by: $v_{i}=s_{i}^{out}+w_{i(N+1)}$.
Now, we introduce a $N \times N$ directed and weighted matrix $M$, such that $m_{ij}=w_{ij}⁄v_i $.
The elements of $M$, represent the one-step transition probability for a single particle (unit of value) to go from $i$ to $j$. Higher powers of M express the transition probability from $i$ to $j$ in a given number of steps. We indicate with $U^{(S)}$, with $S$ integer, the sum of the first $S$ powers of $M$:
\begin{equation}
U^{(S)} \equiv (u^{(S)}_{ij})_{1\le i,j \le N}\equiv  \sum_{q=0}^{S}M^q= \frac{(I-M^{S+1})}{(I-M)}.
\label{eq:matseries}
\end{equation}
Since $M$ is a sub-stochastic matrix, the series converges for  $S \rightarrow \infty$\cite{leontief}. The non-diagonal elements of $U^{(S)}$ represent the probabilities of reaching node $j$ starting from node $i$  within $S$ steps\cite{markov1,markov2}. Therefore, if we take into account all possible paths of any length between two nodes (i.e. $S=\infty$) eq.3 becomes $U^{(\infty)} \equiv (I-M)^{-1}$ . The element $u_{ii}^{(S)}$ enables us to compute the probability to come back to the source node $i$ within $S$ steps. We can therefore compute the cycling index of a vertex as the fraction of trade passing through a node/country $i$ that returns statistically (directly or indirectly) to it within $S$ steps, formally:
\begin{equation}
\gamma_i^{(S)}= \frac{u_{ii}-1}{u_{ii}} \Gamma_i^{(S)} v_i.
\label{eq:singlenode}
\end{equation}
The global cycling index of the network (i.e. the fraction of traded value that returns, directly or indirectly, to a starting node) within $S$ steps is given by: 
\begin{equation}
\Gamma^{(S)}= \frac{\sum_i \gamma_{i}}{\sum_i v_i}= \frac{\sum_i \Gamma_{i}^{(S)} v_i}{\sum_i v_i}
\label{eq:globalnetwork}
\end{equation}
where $\gamma_{i}=\Gamma_{i}^{(S)} v_i$ represents the quantity of cyclic trade (millions of current U.S. dollars), that returns to a node/country $i$ within $S$ steps.
The value of $\Gamma^{(S)}$ ranges between 0 and 1. The former case is observed when there is no trade that starts at some country $i$ that comes back to it. The latter case represent a systems where all trade come back to the starting country. Note that the residual share of trade, up to 1, represent the share of trade that is acyclic, the starting country is different from the ending country.
The quantity $\Gamma^{(\infty)}$ has been used in ecological studies, with the aim of evaluating the total amount of cyclical matter in ecosystems\cite{finn,higashi1,higashi2}.

\subsection{Interpreting the cycling index.}
Ecosystems are open systems exchanging both matter and energy with a source and a sink. The above mentioned cycling index was developed to assess the share of matter that is recycled throughout food chains in an ecological network, from the primary producers (photosynthesis) to the top predators and detritus feeders. Likewise, we can think of added-value as matter in food chains and look by means of cycling how this is conserved throughout the stages of production internationally (GVCs), where the sink and the source of value are national economies. National economies play the role of sink and source in the parallel with ecosystems because the WTW, like ecological networks, is an open system: national economies play the role of sink and source in the parallel with ecosystems because the world trade network, like ecological networks, is an open system: matter is not conserved through every stage of the international production chain, whereby every country is a source of row material and a sink of waste. In this paper the cycling index will be used to assess the GVCs. It is worth noting that the cycling index only measures, statistically, the amount of value that returns to a starting point (country) after a given number of steps (trading relationships), that is, it assesses the cyclical GVCs, from the markets of the raw material, to those of the final products. Three main methodological approaches have been used to capture GVCs in the scientific literature: 1) international trade statistics on parts and components; 2) customs statistics on processing trade and 3) international trade data combined with input-output (I-O) tables. Amador and Cabral provides a detailed review of the existing methodologies\cite{amador14}. More recently, network theory has been applied to disaggregate trade and I-O matrices to investigate GVCs\cite{klaus,shi}. Markov chain theory has been previously applied to disaggregate trade to investigate the allometric scaling of networks and the structure of GVCs\cite{amador15,riccaboni}. Compared to the methods aimed at directly assessing the GVCs by measuring the traded value added in I-O matrices, our approach is different for the following reasons:
1) we do not measure just the share of incorporated value in exports/imports between pairs of countries, i.e. in bilateral relationships, we measure the value that is conserved throughout all possible paths in a network.
2) this is a statistical measure, thus, it does not rely on direct measures of value added (which can only be drawn by I/0 matrices of countries). It has indeed the flaw of being a statistical measure of value, but it has the strength of assessing value through longer-than-one steps of trade (on the contrary, direct measure of value added can only assess one-step path, i.e. between two countries, inasmuch as any following step of the value added this becomes input of production).
3) it is a global (statistical) measure and asses all the possible paths of a given length. In other words, even when assessing two-step cycling, it is statistically relevant all the relationships of i and j with all other countries rather than just the relationship between country i and country j.
4) we statistically asses (with cycling) the value conserved throughout a cycle, for all products, rather than one. This is a statistical measure based on aggregate trade because at every step of the production chain (generally, but not always) products change trade category, i.e. classification: iron, engines, cars, etc.
To use an example from ecological networks - in a food chain, when we want to assess the amount of mass that is conveyed through one species (prey) to the other (predator) at every step of the chain, from primary producers (grass) to the last predators (and decomposers), we cannot tag every atom and check every passage they make. We can only weight body mass of organisms through the food chain. If we know that species A feed 50\% on species B and 50\% on species C, we know that the atoms of the species A  have 0.5 probability of coming from B and 0.5 of coming from C. We can do this for all the species of the food chain and we project this into a continuous, steady food relation. i.e., if species C feeds on species E for 50\%, species A, even if it does not feed on species E, has 0.25 probabilities of having atoms from species E. Upon this, we can calculate the probabilities of an atom to go from one species to the other through all the possible direct and indirect paths. This is referred to as transition matrix, and in the transition matrix, we can calculate the share of atoms that make a cycle, i.e., that start from species A and come back to species A along all the possible paths (not only with species B and C, direct feeding, but also along species E, indirect feeding). Now, suppose we are not talking about atoms, but value of a product. If, for example, Italy sells cars to USA, where the engines of the Italian cars are produced, the share of value of car relative to engine is cyclical with USA. Suppose now that the USA buys iron from China and that Italy sells cars to China. Even if Italy does not buy directly iron from China, the share of the value of iron in the engine of the car is cyclical.

\subsection{Null model as a benchmark}
A popular and appropriate, to our scope, null model is the directed weighted configuration model indeed it preserves the observed intrinsic heterogeneity of vertices: all vertices have the same in-strength and out-strength as in the real network\cite{rec}. In other words, this model preserves the in- and out-strength sequences separately, and, furthermore it preserves the total weight of the original network. To evaluate if our analysis if our analysis is sound and consistent.
We proceed as follows: first, for each year we built the expected network of trade using the randomization approaches of the maximum-likelihood method also called exponential random graph, second, we evaluate $\Gamma^{(S)}$ on the expected networks. See supporting information for the expected trend of $\Gamma^{(S)}$. Note that in doing so we take into account the global trade growth and for each country the distribution of import and export with foreign countries from 1960 to 2011\cite{rec}.

\section{Supporting Information}
\subsection{Methods}

In probability theory a Markov chain is a stochastic process defined on a discrete state space satisfying the Markov property. It is a set of random variables, representing the evolution of a certain system, without memory: each actual state of the system just depends on the previous one. The changes of states are called transitions and the Markov chain can be described by a Markov matrix, $M$, whose elements $m_{ij}$ represent the probabilities from a state $i$ to a state $j$ (transition probabilities or transition rules). Given the memoryless nature of a Markov chain it is not possible to predict the state of a system, in a given time $t$, but it is possible to predict its statistical properties.

In this paper the transition probabilities are set according to the elements of the matrix W that describes the World Trade Web, i.e. the trade network. A link, $w_{ij}$, is assigned to any import/export relationship between two countries, where $w$ is the volume of the trade. In what follow, we explain how the transition rules are assigned.    

In general, given a system of $N$ nodes and a $N \times N$ matrix representing their interactions, the matrix is said balanced if if $s^{out}_i=s^{in}_i,\:\forall\:i \in \{1, \dots , N \}$,  i.e., the sum of each column is equal to the sum of its related row.

Systems described by a balanced matrix can be considered isolated: nodes perfectly balance their total in- and out- flows themselves, without needing any further exchange with the outside. However, most of the network representing real systems are not balanced. This means that $s^{out}_i \neq s^{in}_i$ for at least one $i$ such that $ i \in \{1, \dots , N \}$. In these systems three different sets of nodes can be identified: the set of vertices with $s^{out}_i=s^{in}_i$, the set of vertices with $s^{out}_i>s^{in}_i$ and the set of vertices with $s^{out}_i<s^{in}_i$.

In the first group, the total ingoing and outgoing fluxes of all nodes are balanced. In the second set, each vertex needs an extra incoming weight in order to balance in and out strengths. In this case, in order to balance the nodes in this set, we introduce an additional node/vertex, called \emph{source} and labelled with $0$, providing extra-ingoing fluxes to all nodes in the set. The new link between the source and each node in the set will have the following weight: $w_{0i}=s^{out}_i - s^{in}_{i}$. Similarly, a new vertex is introduced for the nodes in the third group in order to balance their in-strengths. It is called \emph{sink} and labelled $N+1$. The new links between this latter and the nodes in the set will have weights equal to: $w_{i (N+1)}=s^{in}_i-s^{out}_{i}$.

For each node $i$ with $i \in \{1,\dots,N \}$ the \emph{total outgoing flow}, $v_{i}$, is given by: $v_{i}=s_{i}^{out}+w_{i(N+1)}$. Now, we introduce now a $N \times N$ directed and weighted matrix $M$, such that  $m_{ij}=w_{ij}/v_{i}$. 
We indicate with $U^{(S)}$, with $S$ integer, the sum of the first $S$  powers of $M$: 
\begin{equation} \begin{split}
U^{(S)} \equiv (u^{(S)}_{ij})_{1\le i,j \le N}\equiv I + M^1 + M^2 +  \dots + M^S\equiv \\
\equiv \sum_{q=0}^{S}M^q= \frac{(I-M^{S+1})}{(I-M)}.
\end{split}
\end{equation}
Since M is a sub-stochastic matrix, the series converges  for $S \rightarrow \infty$ \cite{leontief}.

The diagonal elements of $U^{(S)}$ allow the evaluation of the cycling index, $\Gamma_i^{(S)}$; for each country according to the following $\Gamma_i^{(S)}= \frac{u^{(S)}_{ii}-1}{u^{(S)}_{ii}}$. It is $\Gamma_i^{(S)}$ is a fraction of $v_i$, therefore we can evaluate the total value of the flow (directly or indirectly) returning to a node as:
\begin{equation}
\gamma_i^{(S)}= \Gamma_i^{(S)} v_i
\label{eq:singlenode}
\end{equation}

The global cycling index within $S$ steps is given by $\Gamma^{(S)}= (\sum_i \Gamma_{i}^{(S)} v_i)/(\sum_i v_i)$.

\subsubsection{Two extreme cases}
The quantity $\Gamma^{(S)}$ has a lower bound, 0, and upper bound, 1. The two extreme cases are $\Gamma^{(S)} = 0$ and $ \lim\limits_{S \to \infty} \Gamma^{(S)} = 1$. The former is observed in \emph{Directed Acyclic Graphs} (DAG), and the latter in balanced networks. While the case $\Gamma^{(S)} =0$ is straightforward due to the fact that circular paths do not exist in a DAG, the second case is not trivial to picture. We give a quite intuitive idea of the proof in two steps.

Let us assume a balanced and connected network.  In network theory, two nodes $i$ and $j$ are \emph{strongly connected} if there exists a path from $i$ to $j$ and vice-versa. This definition of the relation is reflexive, symmetric and transitive, dividing nodes in disjoint sets of equivalence.  Each class is defined SCC and among them all nodes are strongly connected to each other, but no bidirectional path exists between the classes.

First, we will prove by contradiction that a balanced and connected network can have one and only one SCC, i.e., each node is reachable from any other node. Thus, let us assume that we have a network, represented by the matrix $W$, with two distinct SCCs (see picture).


We suppose that $p$ of the $N$ nodes of the networks belong to $SCC_1$, while the remaining $z$ vertices belong to $SCC_2$. The network is connected, thus exists at least a link $w_{ik}$ from a node $i \in SCC_1$ to a node $k \in SCC_2$.  

Since that the network is balanced we can write the sum of each row of matrix W  equal to the sum of the related column. We write down this equality for the nodes in $SCC_1$, and we explicit the weights related to the node i:

\begin{eqnarray*}
 \sum_{j=1}^p w_{1j}+ w_{1i} & = & \sum_{j=1}^p w_{j1} + w_{i1} \\
 \sum_{j=1}^p w_{2j}+ w_{2i} & = & \sum_{j=1}^p w_{j2} + w_{i2}  \\
& \dots &  \\
 \sum_{j=1}^p w_{pj}+ w_{pi} & = & \sum_{j=1}^p w_{jp} + w_{ip}.
\end{eqnarray*}

If we sum up all the left side members and right side members in the equalities and simplify, we obtain: $ \sum_j^p w_{ji}= \sum_j^p w_{ij}$. We observe that $\sum_j^p w_{ji} =s_i^{in} $, while $\sum_j^p w_{ij}= s_i^{out} - w_{ik}$. The network is balanced by hypothesis, therefore $w_{ik}=0$. This result contradicts the assumption that the network is connected, leading to the conclusion that it is impossible to have more than one SCC in a balanced and connected network.

\begin{figure}[]
\subfigure[]
{
\label{fig:USA}
\includegraphics[width=0.45\textwidth]{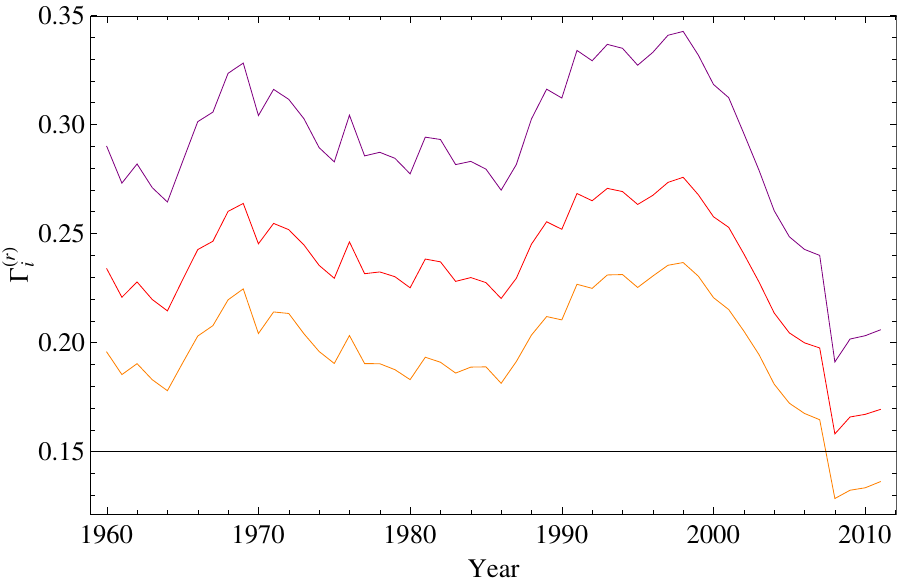}
}
\subfigure[]
{
\label{fig:Germany}
\includegraphics[width=0.45\textwidth]{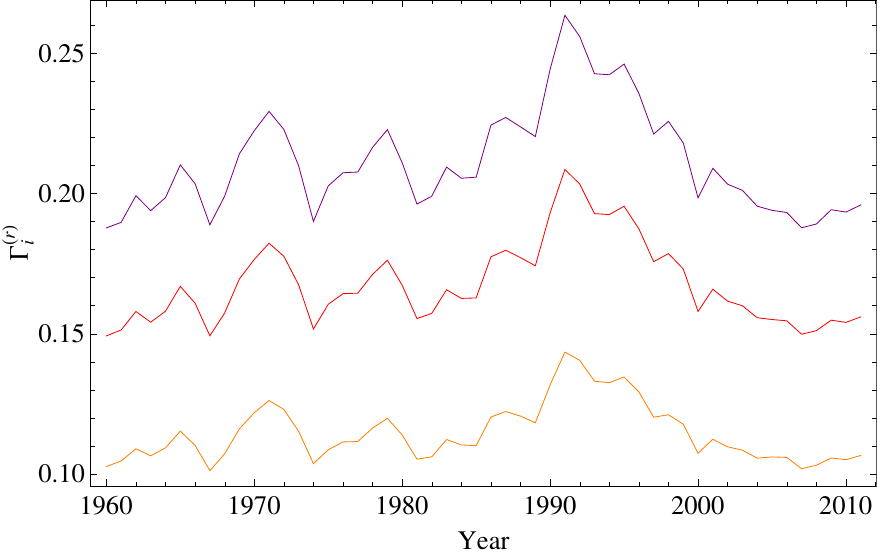}
}
\caption{From 1960 to 2012, the trends of $\Gamma_i^{(2)}$ (orange line), $\Gamma_i^{(3)}$ (red line) and $\Gamma_i^{(4)}$ (purple line) for the USA (panel a) and Germany (panel b).}
\label{fig:stepsex}

\end{figure}

The second step of the proof consists in showing that for a balanced network $u_{ii}^{(S)} \rightarrow \infty $ for $S \rightarrow \infty$.
We know that a balanced and connected network is strongly connected and implies that the related Markov chain is {\emph irreducible}, in other terms it is a single communicating class. A state $i$ of a Markov chain is called recurrent if the probability to start from $i$ and come back to it is equal to 1. In particular, if the chain is finite and irreducible, then all states are recurrent \cite{markov}.

In an irreducible Markov chain with finite number of nodes the nodes are recurrent. Nodes are recurrent if and only if $\sum_S^{\infty} u_{ii}^{(S)}=\infty$, hence eq. \ref{eq:singlenode} becomes: $\Gamma^{(\infty)}= u_{ii}^{(\infty)}-1/u_{ii}^{(\infty)}=1$. We conclude that for a balanced connected network $\Gamma^{(\infty)}=1$.

\begin{table}
\centering
\begin{tabular}{| c || c | c | c | c |}
\hline

Country  & $ \Gamma_i^{(2)}$ & $\Gamma_i^{(3)}$ & $ \Gamma_i^{(4)}$ &  $\Gamma_i^{(\infty)}$   \\ \hline\hline
CHI (1961) & 0.03(0.8) & $ 0.08(0.5)$ &0.10(0.4) & 0.18(0.2)\\  \hline   
FRA (1961) & 0.63 & 0.53  & 0.47  & 0.18 (0.2)\\  \hline
GER (1971) & 0.25(0.1) & 0.26(0.1) & 0.26(0.1) & 0.26(0.1)\\  \hline
JAP (1961) & 0.76   & 0.75 & 0.75 & 0.44  \\  \hline
U.K. (1961) &0.16($0.2$) &0.19($0.2$) & 0.2($ 0.1 $) & 0.22(0.1)\\  \hline
RUS (1990) &-0.57($0.02$) &-0.54 (0.01) &-0.53($0.01 $) &-0.54(0.01)\\  \hline
USA (1961) & 0.31(0.02) & 0.31(0.03) & 0.31(0.03) & 0.28(0.04) \\  \hline

\end{tabular}
\caption{For different countries, the table reports the Pearson's correlation index between values of $\Gamma_i^{(r)}$ (for $r=2,3,4,\infty$) and GDP growth at constant price, in parenthesis the p-value ($95\%$ confidence, and p-value lower than or equal to $10^3$ are not reported) . GDP growth data are from World Bank; the starting year of World Bank statistic is reported near each country.}
    
\label{tab:corr}
\end{table}

\subsection{Supplementary Discussion}

The method described to build $M$ has been used in ecological studies, with the aim of evaluating the total amount of matter cycling in the ecosystem with the quantity $\Gamma^{(\infty)}$ \cite{Finn,Higashi,Higashi1}. Recently this approach has been employed to detect the presence of allometry patterns both in ecosystems and World Trade Web (WTW)\cite{Zhang,allometryWTW}. The novelty of our methodology stands in considering all possible orders of cycling paths, with S to $\{2,...,\infty\}$, in order to evaluate the Global Chain Value, and the share of cyclic trade in the world economy.

In this perspective, the three different sets of nodes defined can be thought as groups of countries with a perfect balance between export and import ($s^{out}_i=s^{in}_i$), a surplus of exports $s^{out}_i>s^{in}_i$ or of imports $s^{out}_i<s^{in}_i$. According to this interpretation, the sink and source nodes just represent the total of the World's national economies.

In the paper we showed the strong correlation between the global cycling index and the crude oil price. This is the main result of our work in terms of policy implications, nevertheless the single country analysis reveals interesting information.         

The two quantities $\gamma_i^{(S)}$ and $\Gamma_i^{(S)}$ can be used to evaluate the performances of each country from 1960 to 2011. Their trends can reveal significant differences in the structure of the chain values and shed light onto the distances in the underlying economic system.

Figure 4 shows the fluctuations of a sample of $\Gamma_i^{(S)}$  for USA (left panel) and Germany (right panel). 

For both countries, a big share of exports returns directly or indirectly to the country, however the plots depict different cycling patterns. Specifically, taking into account ciclyc paths longer than 2 steps, USA seems characterized by a longer chain of production then Germany. Indeed, the share of USA cyclic paths equal to 4 steps is bigger than the 3 steps one, while the majority of German cyclic paths show length equal to 3. 
Furthermore the trends of $\Gamma_i$ curves show a different response of the two countries to the 2008 crisis. A decrease of the share of cyclic flows is observable since the end of 90s and it culminates into a dramatic drop in 2008.

The overall percentage variation is around $-15\%$. On the other hand, the German declining trend starts around 1990 and seems to stabilize from 2000 on, and the 2008 crisis appears negligible. The overall percentage variation of the share of cycling paths is around $-5\%$.

We could argue that the cycling coefficients and the growth of Gross Domestic Product of a country are related. However this connection strongly varies from country to country without a clear pattern. Indeed, in table (II) we show the Pearson's correlation coefficient between the values of GDP growth (p-values in brackets) at constant price  and $ \Gamma_i^{(S)}$, with $S \in \{2,3,4, \infty \}$, for six countries: China, France, Germany, Japan, United Kingdom, Russia and USA (GDP data comes from World Bank, www.worldbank.org).  
Most of the countries do not show any significant correlation or just a slight one; France and Japan exhibit a strong positive correlation between the two variables, while for Russia we found a quite high negative correlation. It means that the information captured by gamma substantially differs from what can be drawn from the analysis of the GDP growth.

\begin{figure*}
\begin{center}

\subfigure[]
{
\label{fig:RankAbs}
\includegraphics[width=0.48\textwidth]{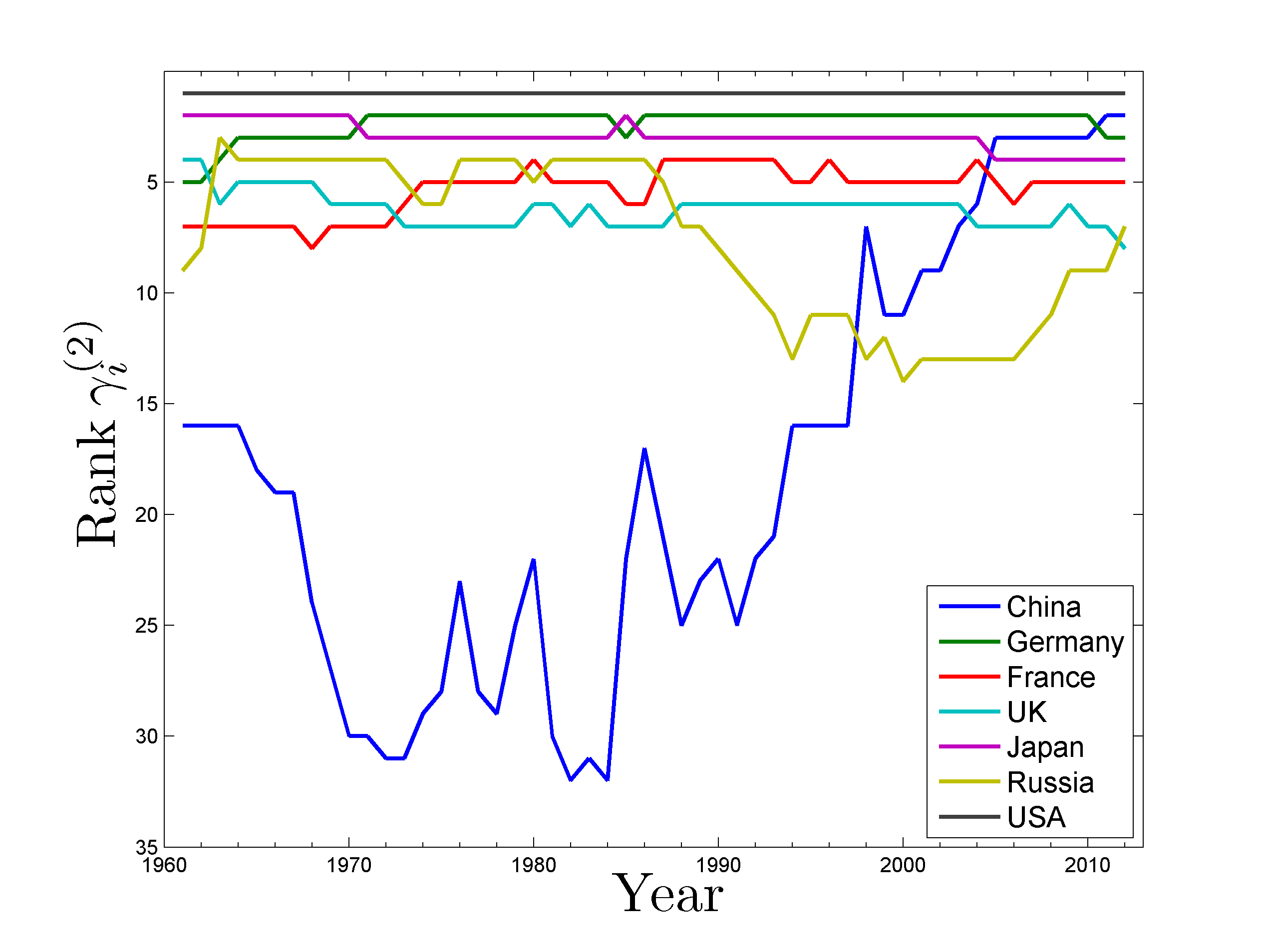}
}
\subfigure[]
{
\label{fig:RankAbs}
\includegraphics[width=0.48\textwidth]{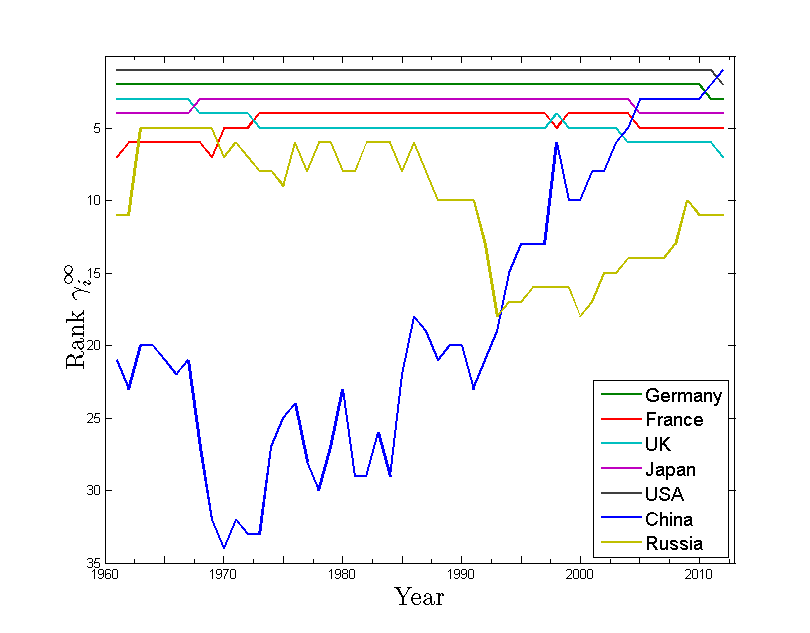}
}
\subfigure[]
{
\label{fig:4stepsA}
\includegraphics[width=0.48\textwidth]{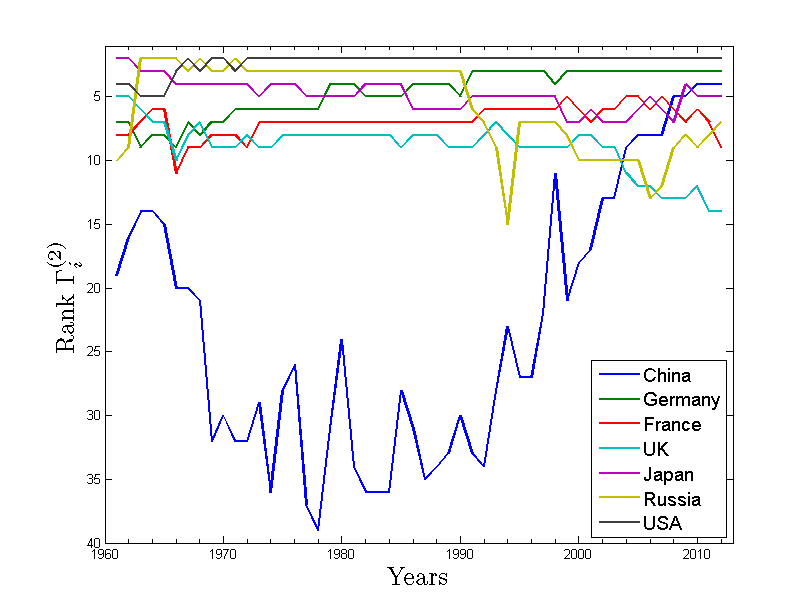}
}
\subfigure[]
{
\label{fig:4stepsB}
\includegraphics[width=0.48\textwidth]{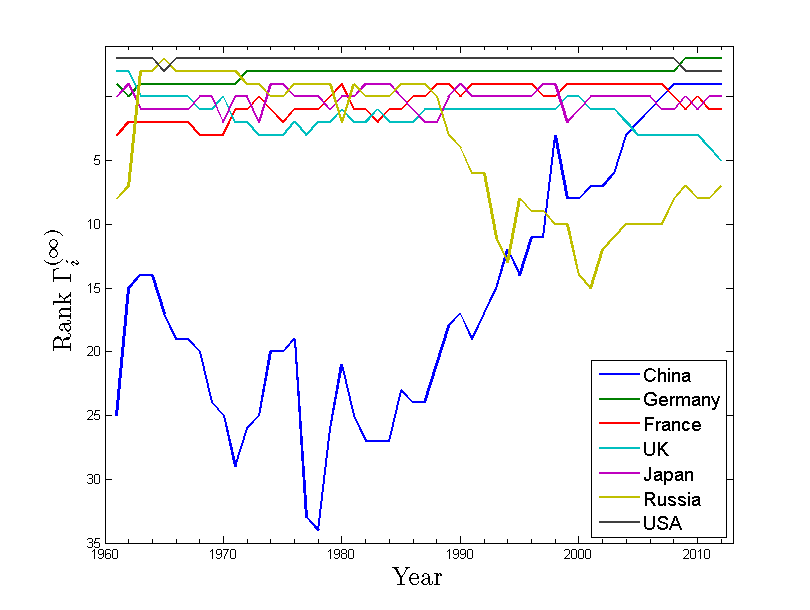}
}
\\ 
\end{center}
    \caption{The ranking position for seven countries according to the value of $\gamma_i^{(2)}$ (panel a), $\gamma_i^{(\infty)}$ (panel b), $\Gamma_i^{(2)}$ (panel c) and $\Gamma_i^{(\infty)}$ (panel d). The data range covers from 1960 to 2011.}
\label{fig:stepsex}
\end{figure*}

The $\gamma_i^{(S)}$ and $\Gamma_i^{(S)}$ can give information about the performance of single countries from 1960 to 2011. 
It is possible to rank countries according to their values. In figure (5) we show this rank, for $S \in \{ 2, inf \}$, of the same seven countries: China, Germany, France, UK, Japan, Russia, and USA.

We chose these countries for their interesting patterns in order to reveal how the variables $\gamma_i^{(S)}$ and $\Gamma_i^{(S)}$  can explain structural changes in countries. Russia position is always higher in the relative than in the absolute rank, and for $S=2$ than for $S=\infty$. Moreover we can observe a decline in both starting in late 80s and a rise since around 2000. Results seem in line with the USSR dissolution, explained by a Russian chain value made up substantially by bilateral exchanges with the other Soviet Republics.

China is characterized by a remarkable increasing trend in the last decades, reaching the top of the rank for $\gamma_i^{(\infty)}$ in the last years of our analysis. In the same period, Germany overtakes USA share of cyclic trade ($S=\infty$), as emerges from fig. 5 panel d.

We remark on the performance of China and Germany. The former has been on the rise in the last decades and we point out the pervasiveness of its chain value in absolute value ($\gamma_i^{(\infty)}$, fig. 5 panel b). In the last year regarding $S=\infty$, Germany has outdone USA for share of cyclic trade (fig. 5 panel d).

\bibliographystyle{plainnat}
\bibliography{scibib}

\begin{thebibliography}{52}
\providecommand{\natexlab}[1]{#1}
\providecommand{\url}[1]{\texttt{#1}}
\expandafter\ifx\csname urlstyle\endcsname\relax
  \providecommand{\doi}[1]{doi: #1}\else
  \providecommand{\doi}{doi: \begingroup \urlstyle{rm}\Url}\fi

\bibitem[Allsopp and Fattouh(2011)]{allsopp}
Christopher Allsopp and Bassam Fattouh.
\newblock Oil and international energy.
\newblock \emph{Oxford Review of Economic Policy}, 27\penalty0 (1):\penalty0
  1--32, 2011.

\bibitem[Alvarez-Ramirez et~al.(2010)Alvarez-Ramirez, Alvarez, and
  Solis]{alvarez}
Jose Alvarez-Ramirez, Jesus Alvarez, and Ricardo Solis.
\newblock Crude oil market efficiency and modeling: Insights from the
  multiscaling autocorrelation pattern.
\newblock \emph{Energy Economics}, 32\penalty0 (5):\penalty0 993--1000, 2010.

\bibitem[Amador and Cabral(2014)]{amador14}
Jo{\~a}o Amador and S{\'o}nia Cabral.
\newblock Global value chains: a survey of drivers and measures.
\newblock \emph{Journal of Economic Surveys}, 2014.

\bibitem[Amador et~al.(2015)Amador, Cabral, et~al.]{amador15}
Jo{\~a}o Amador, S{\'o}nia Cabral, et~al.
\newblock Networks of value added trade.
\newblock Technical report, 2015.

\bibitem[Anderson and Van~Wincoop(2004)]{anderson04}
James~E Anderson and Eric Van~Wincoop.
\newblock Trade costs.
\newblock Technical report, National Bureau of Economic Research, 2004.

\bibitem[Ayres et~al.(2003)Ayres, Ayres, and Warr]{ayres}
Robert~U Ayres, Leslie~W Ayres, and Benjamin Warr.
\newblock Exergy, power and work in the us economy, 1900--1998.
\newblock \emph{Energy}, 28\penalty0 (3):\penalty0 219--273, 2003.

\bibitem[Baldwin(2006)]{baldwin}
Richard Baldwin.
\newblock Globalisation: the great unbundling (s).
\newblock \emph{Economic Council of Finland}, 20\penalty0 (2006):\penalty0
  5--47, 2006.

\bibitem[Banavar et~al.(1999)Banavar, Maritan, and Rinaldo]{banavar}
Jayanth~R Banavar, Amos Maritan, and Andrea Rinaldo.
\newblock Size and form in efficient transportation networks.
\newblock \emph{Nature}, 399\penalty0 (6732):\penalty0 130--132, 1999.

\bibitem[Barsky and Kilian(2004)]{barsky}
Robert Barsky and Lutz Kilian.
\newblock Oil and the macroeconomy since the 1970s.
\newblock Technical report, National Bureau of Economic Research, 2004.

\bibitem[Bharucha-Reid(2012)]{markov2}
Albert~T Bharucha-Reid.
\newblock \emph{Elements of the Theory of Markov Processes and their
  Applications}.
\newblock Courier Dover Publications, 2012.

\bibitem[{British Petroleum}(2010)]{oil}
Global {British Petroleum}.
\newblock {BP statistical review of world energy}, June 2010.

\bibitem[Carrasco and Serrano(2014)]{carrasco}
Carlos~A Carrasco and Felipe Serrano.
\newblock Global and european imbalances: A critical review.
\newblock Technical report, 2014.

\bibitem[Coe et~al.(2007)Coe, Subramanian, and Tamirisa]{coe07}
David~T Coe, Arvind Subramanian, and Natalia~T Tamirisa.
\newblock The missing globalization puzzle: Evidence of the declining
  importance of distance.
\newblock \emph{IMF staff papers}, pages 34--58, 2007.

\bibitem[Cologni and Manera(2009)]{cologni}
Alessandro Cologni and Matteo Manera.
\newblock The asymmetric effects of oil shocks on output growth: A
  markov--switching analysis for the g-7 countries.
\newblock \emph{Economic Modelling}, 26\penalty0 (1):\penalty0 1--29, 2009.

\bibitem[Feenstra and Hanson(1997)]{feenstra97}
Robert~C Feenstra and Gordon~H Hanson.
\newblock Productivity measurement and the impact of trade and technology on
  wages: Estimates for the us, 1972-1990.
\newblock Technical report, National Bureau of Economic Research, 1997.

\bibitem[Finn(1976)]{finn}
John~T. Finn.
\newblock Measures of ecosystem structure and function derived from analysis of
  flows.
\newblock \emph{Journal of Theoretical Biology}, 56\penalty0 (2):\penalty0 363
  -- 380, 1976.

\bibitem[Garlaschelli and Loffredo(2005)]{diego05}
Diego Garlaschelli and Maria~I Loffredo.
\newblock {Structure and evolution of the world trade network}.
\newblock \emph{Physica A: Statistical Mechanics and its Applications},
  355\penalty0 (1):\penalty0 138--144, 2005.

\bibitem[Gaulier and Zignago(2010)]{zignago}
Guillaume Gaulier and Soledad Zignago.
\newblock Baci: international trade database at the product-level (the
  1994-2007 version).
\newblock 2010.

\bibitem[Gleditsch(2002)]{gled}
Kristian~Skrede Gleditsch.
\newblock Expanded trade and gdp data.
\newblock \emph{Journal of Conflict Resolution}, 46\penalty0 (5):\penalty0
  712--724, 2002.

\bibitem[Granger(1969)]{granger}
Clive~WJ Granger.
\newblock Investigating causal relations by econometric models and
  cross-spectral methods.
\newblock \emph{Econometrica: Journal of the Econometric Society}, pages
  424--438, 1969.

\bibitem[Hamilton(1983)]{hamilton}
James~D Hamilton.
\newblock Oil and the macroeconomy since world war ii.
\newblock \emph{The Journal of Political Economy}, pages 228--248, 1983.

\bibitem[Hamilton(2011)]{hamilton11}
James~D Hamilton.
\newblock Nonlinearities and the macroeconomic effects of oil prices.
\newblock \emph{Macroeconomic dynamics}, 15\penalty0 (S3):\penalty0 364--378,
  2011.

\bibitem[Higashi et~al.(1993)Higashi, Patten, and Burns]{higashi2}
Masahiko Higashi, Bernard~C. Patten, and Thomas~P. Burns.
\newblock Network trophic dynamics: the modes of energyutilization in
  ecosystems.
\newblock \emph{Ecological Modelling}, 66:\penalty0 1--42, 1993.

\bibitem[Hooker(1996)]{hooker}
Mark~A Hooker.
\newblock What happened to the oil price-macroeconomy relationship?
\newblock \emph{Journal of monetary Economics}, 38\penalty0 (2):\penalty0
  195--213, 1996.

\bibitem[Hummels(2007)]{hummels07}
David Hummels.
\newblock Transportation costs and international trade in the second era of
  globalization.
\newblock \emph{The Journal of Economic Perspectives}, pages 131--154, 2007.

\bibitem[Hummels et~al.(2001)Hummels, Ishii, and Yi]{hummels01}
David Hummels, Jun Ishii, and Kei-Mu Yi.
\newblock The nature and growth of vertical specialization in world trade.
\newblock \emph{Journal of international Economics}, 54\penalty0 (1):\penalty0
  75--96, 2001.

\bibitem[Jones and Leiby(1996)]{jones}
Donald~W Jones and Paul~N Leiby.
\newblock The macroeconomic impacts of oil price shocks: A review of literature
  and issues.
\newblock \emph{Oak Ridge National Laboratory}, 1996.

\bibitem[Jones et~al.(2004)Jones, Leiby, and Paik]{jones04}
Donald~W Jones, Paul~N Leiby, and Inja~K Paik.
\newblock Oil price shocks and the macroeconomy: what has been learned since
  1996.
\newblock \emph{The Energy Journal}, pages 1--32, 2004.

\bibitem[Levinson and Taylor(2008)]{levinson}
Arik Levinson and M~Scott Taylor.
\newblock Unmasking the pollution haven effect*.
\newblock \emph{International economic review}, 49\penalty0 (1):\penalty0
  223--254, 2008.

\bibitem[Milo et~al.(2002)Milo, Shen-Orr, Itzkovitz, Kashtan, Chklovskii, and
  Alon]{milo}
Ron Milo, Shai Shen-Orr, Shalev Itzkovitz, Nadav Kashtan, Dmitri Chklovskii,
  and Uri Alon.
\newblock Network motifs: simple building blocks of complex networks.
\newblock \emph{Science}, 298\penalty0 (5594):\penalty0 824--827, 2002.

\bibitem[Miroudot et~al.(2009)Miroudot, Lanz, and Ragoussis]{miroudot}
S{\'e}bastien Miroudot, Rainer Lanz, and Alexandros Ragoussis.
\newblock Trade in intermediate goods and services.
\newblock 2009.

\bibitem[Mork(1989)]{mork}
Knut~Anton Mork.
\newblock Oil and the macroeconomy when prices go up and down: an extension of
  hamilton's results.
\newblock \emph{Journal of political Economy}, pages 740--744, 1989.

\bibitem[Naccache(2010)]{naccache}
Th{\'e}o Naccache.
\newblock Slow oil shocks and the “weakening of the oil price--macroeconomy
  relationship”.
\newblock \emph{Energy Policy}, 38\penalty0 (5):\penalty0 2340--2345, 2010.

\bibitem[Norris(1998)]{markov1}
James~R Norris.
\newblock \emph{Markov chains}.
\newblock Number 2008. Cambridge university press, 1998.

\bibitem[Oladosu(2009)]{ola}
Gbadebo Oladosu.
\newblock Identifying the oil price-macroeconomy relationship: An empirical
  mode decomposition analysis of us data.
\newblock \emph{Energy Policy}, 37\penalty0 (12):\penalty0 5417--5426, 2009.

\bibitem[Papapetrou(2001)]{papa}
Evangelia Papapetrou.
\newblock Oil price shocks, stock market, economic activity and employment in
  greece.
\newblock \emph{Energy Economics}, 23\penalty0 (5):\penalty0 511--532, 2001.

\bibitem[Park and Ratti(2008)]{park08oil}
Jungwook Park and Ronald~A Ratti.
\newblock Oil price shocks and stock markets in the us and 13 european
  countries.
\newblock \emph{Energy Economics}, 30\penalty0 (5):\penalty0 2587--2608, 2008.

\bibitem[Patten and Higashi(1995)]{higashi1}
Bernard~C. Patten and Masahiko Higashi.
\newblock First passage flows in ecological networks: measurement by
  input-output flow analysis.
\newblock \emph{Ecological Modelling}, 79\penalty0 (1–3):\penalty0 67 -- 74,
  1995.

\bibitem[Prell et~al.(2014)Prell, Feng, Sun, Geores, and Hubacek]{klaus}
Christina Prell, Kuishuang Feng, Laixiang Sun, Martha Geores, and Klaus
  Hubacek.
\newblock The economic gains and environmental losses of us consumption: A
  world-systems and input-output approach.
\newblock \emph{Social Forces}, 93\penalty0 (1):\penalty0 405--428, 2014.

\bibitem[Rebucci and Spatafora(2006)]{rebucci}
Alessandro Rebucci and Nikola Spatafora.
\newblock Oil prices and global imbalances.
\newblock \emph{IMF World Economic Outlook}, 4\penalty0 (2006):\penalty0
  71--96, 2006.

\bibitem[Ruzzenenti and Basosi(2008)]{francorebound}
Franco Ruzzenenti and Riccardo Basosi.
\newblock The rebound effect: An evolutionary perspective.
\newblock \emph{Ecological Economics}, 67\penalty0 (4):\penalty0 526--537,
  2008.

\bibitem[Ruzzenenti et~al.(2010)Ruzzenenti, Garlaschelli, and
  Basosi]{francosymmetryII}
Franco Ruzzenenti, Diego Garlaschelli, and Riccardo Basosi.
\newblock Complex networks and symmetry ii: reciprocity and evolution of world
  trade.
\newblock \emph{Symmetry}, 2\penalty0 (3):\penalty0 1710--1744, 2010.

\bibitem[Shi et~al.(2014)Shi, Zhang, Yang, and Luo]{shi}
Peiteng Shi, Jiang Zhang, Bo~Yang, and Jingfei Luo.
\newblock Hierarchicality of trade flow networks reveals complexity of
  products.
\newblock \emph{PloS one}, 9\penalty0 (6):\penalty0 e98247, 2014.

\bibitem[Solow(1952)]{leontief}
Robert Solow.
\newblock On the structure of linear models.
\newblock \emph{Econometrica: Journal of the Econometric Society}, pages
  29--46, 1952.

\bibitem[Squartini and Garlaschelli(2012)]{motifs}
Tiziano Squartini and Diego Garlaschelli.
\newblock Triadic motifs and dyadic self-organization in the world trade
  network.
\newblock In \emph{Self-Organizing Systems}, pages 24--35. Springer, 2012.

\bibitem[Squartini et~al.(2013)Squartini, Picciolo, Ruzzenenti, and
  Garlaschelli]{rec}
Tiziano Squartini, Francesco Picciolo, Franco Ruzzenenti, and Diego
  Garlaschelli.
\newblock {Reciprocity of weighted networks}.
\newblock \emph{Scientific reports}, 3, 2013.

\bibitem[Stouffer et~al.(2007)Stouffer, Camacho, Jiang, and Amaral]{camacho}
Daniel~B Stouffer, Juan Camacho, Wenxin Jiang, and Lu{\'i}s A~Nunes Amaral.
\newblock Evidence for the existence of a robust pattern of prey selection in
  food webs.
\newblock \emph{Proceedings of the Royal Society B: Biological Sciences},
  274\penalty0 (1621):\penalty0 1931--1940, 2007.

\bibitem[Transport(2009)]{transport}
Energy Transport.
\newblock Co2: Moving towards sustainability.
\newblock \emph{International Energy Agency}, page~44, 2009.

\bibitem[Zamora-L{\'o}pez et~al.(2008)Zamora-L{\'o}pez, Zlati{\'c}, Zhou,
  \v{S}tefan\v{c}i{\'c}, and Kurths]{zamora}
Gorka Zamora-L{\'o}pez, Vinko Zlati{\'c}, Changsong Zhou, Hrvoje
  \v{S}tefan\v{c}i{\'c}, and J{\"u}rgen Kurths.
\newblock Reciprocity of networks with degree correlations and arbitrary degree
  sequences.
\newblock \emph{Physical Review E}, 77\penalty0 (1):\penalty0 016106, 2008.

\bibitem[Zhu et~al.(2015)Zhu, Puliga, Cerina, Chessa, and Riccaboni]{riccaboni}
Zhen Zhu, Michelangelo Puliga, Federica Cerina, Alessandro Chessa, and Massimo
  Riccaboni.
\newblock Global value trees.
\newblock 2015.

\bibitem[Zlati{\'c} and \v{S}tefan\v{c}i{\'c}(2009)]{vinko09}
Vinko Zlati{\'c} and Hrvoje \v{S}tefan\v{c}i{\'c}.
\newblock Influence of reciprocal edges on degree distribution and degree
  correlations.
\newblock \emph{Physical Review E}, 80\penalty0 (1):\penalty0 016117, 2009.

\bibitem[Zlati{\'c} and \v{S}tefan\v{c}i{\'c}(2011)]{vinko11}
Vinko Zlati{\'c} and Hrvoje \v{S}tefan\v{c}i{\'c}.
\newblock {Model of wikipedia growth based on information exchange via
  reciprocal arcs}.
\newblock \emph{EPL (Europhysics Letters)}, 93\penalty0 (5):\penalty0 58005,
  2011.

\end{thebibliography}

\end{document}